\documentclass[%
 reprint,
superscriptaddress,
 amsmath,amssymb,
 aps,
floatfix,
]{revtex4-2}

\usepackage[caption=false]{subfig}
\usepackage{graphicx}% Include figure files
\usepackage[outdir=./]{epstopdf}
\usepackage{dcolumn}% Align table columns on decimal point
\usepackage{bm}% bold math
\usepackage{hyperref}
\usepackage[hyphenbreaks]{breakurl}
\begin{document}

\preprint{APS/123-QED}

\title{A kinetic interpretation of the classical Rayleigh-Taylor instability}% Force line breaks with \\

\author{John Rodman}
\affiliation{Virginia Tech, Blacksburg, Virginia 24061}

\author{Petr Cagas}
\affiliation{Virginia Tech, Blacksburg, Virginia 24061}

\author{Ammar Hakim}
\affiliation{Princeton Plasma Physics Laboratory, Princeton, New Jersey 08543}

\author{Bhuvana Srinivasan}
\email{srinbhu@vt.edu}
\affiliation{Virginia Tech, Blacksburg, Virginia 24061}

\date{\today}% It is always \today, today,
             %  but any date may be explicitly specified

\begin{abstract}
Rayleigh-Taylor (RT) instabilities are prevalent in many physical regimes ranging from astrophysical to laboratory plasmas and have primarily been studied using fluid models, the majority of which have been ideal fluid models. This work is the first of its kind to present a 5-dimensional (2 spatial dimensions, 3 velocity space dimensions) simulation using the continuum-kinetic model to study the effect of the collisional mean-free-path and transport on the instability growth. The continuum-kinetic model provides noise-free access to the full particle distribution function permitting a detailed investigation of the role of kinetic physics in hydrodynamic phenomena such as the RT instability. For long mean-free-path, there is no RT instability growth, but as collisionality increases, particles relax towards the Maxwellian velocity distribution, and the kinetic simulations reproduce the fluid simulation results. An important and novel contribution of this work is in the intermediate collisional cases that are not accessible with traditional fluid models and require kinetic modeling.  Simulations of intermediate collisional cases show that the RT instability evolution is significantly altered compared to the highly collisional fluid-like cases.  Specifically, the growth rate of the intermediate collisionality RT instability is lower than the high collisionality case while also producing a significantly more diffused interface. The higher moments of the distribution function play a more significant role relative to inertial terms for intermediate collisionality during the evolution of the RT instability interface. Particle energy-flux is calculated from moments of the distribution and shows that transport is significantly altered in the intermediate collisional case and deviates much more so from the high collisionality limit of the fluid regime.
\end{abstract}

%\keywords{Suggested keywords}%Use showkeys class option if keyword
                              %display desired
\maketitle

%\tableofcontents

\section{\label{sec:intro}Introduction}

Rayleigh-Taylor (RT) instabilities occur when a dense fluid is accelerated into a lighter fluid, for example under the influence of a gravitational field \citep{Taylor1950, Rayleigh1882}. While this instability is traditionally studied in a strictly fluid regime \citep{Sharp1984, Kull1991}, applying a fully-kinetic treatment allows for study of a range of collisionality, from collisionless and intermediate, where fluid models are not applicable, to highly collisional regimes approaching the fluid limit \citep{ramaprabhu2006RT,Sagert2015, Gallis2016,Wei2012}. 

This work explores a fully-kinetic treatment of the classical RT instability for a single neutral particle species for varying collisionality, with a future goal of extending into a collisional two-species plasma with evolving electromagnetic fields.  A body of literature exists studying magnetohydrodynamic and extended-magnetohydrodynamic modeling of the RT instability \citep{stone2007magnetic, srinivasan2012mechanism, srinivasan2012magnetic}, the role of viscosity, resistivity, and thermal conduction in RT and magneto-RT instability growth \citep{srinivasan2014mitigating, song2020survey, bera2021effect}, and the role of incorporating some kinetic effects on the magneto-RT instability through use of higher-fidelity fluid models \citep{Huba1998, srinivasan2018role}.  

Kinetic effects can emerge when mean-free-paths are long relative to a relevant characteristic length scale. Shock-driven implosion experiments at the OMEGA Laser facility \citep{Omega} have shown evidence of kinetic phenomena in high-energy-density regimes, such as non-hydrodynamic mixing, thermal decoupling, and species separation \citep{Rinderknecht2014, Rinderknecht2015, Rosenberg2014}. Emergence of kinetic effects within a shock may imply the presence of kinetic effects for the RT instability when mean-free-paths are long relative to the fluid interface. Other implosion experiments at OMEGA have studied the physics relevant to RT instability growth in core-collapse supernovae but focused on a purely hydrodynamic interpretation of the results \citep{KuranzDrake2009,Kuranz2009,Drake2004}. As there is evidence of a transition from a hydrodynamic to a kinetic regime within OMEGA high-energy-density experiments, fully-kinetic simulations to accompany RT experiments may offer a novel explanation of disparities between experiment and hydrodynamic simulation.

For these studies, the continuum-kinetic capabilities of the plasma simulation framework \texttt{Gkeyll} \citep{gkylDocs} are used to evolve particle distribution functions, $f$. \texttt{Gkeyll} uses a discontinuous Galerkin method \citep{Reed1973, Cockburn1998, Cockburn2001} to discretize and evolve the Boltzmann equation \citep{Juno2018, Hakim2020},

\begin{equation}
	\frac{\partial{}f}{\partial t} +  \boldsymbol{v}\cdot\boldsymbol{\nabla}_{\boldsymbol{x}} f + \boldsymbol{a}\cdot\boldsymbol{\nabla}_{\boldsymbol{v}}f = \left( \frac{\partial{f}}{\partial t}  \right)_C,
\end{equation}
where $\boldsymbol{x}$ and $\boldsymbol{v}$ are the two independent particle position and velocity, respectively. Acceleration vector, $\boldsymbol{a}$, is simply gravity, $\boldsymbol{g}$, for this work, as only neutral particles will be considered. The right-hand term accounts for particle collisions and is approximated here by the Bhatnagar-Gross-Krook (BGK) operator \citep{Bhatnagar1951, Cagas2018}, 

\begin{equation}
	\left( \frac{\partial{f}}{\partial t}  \right)_C = \nu(f_{M} - f),
\end{equation}
where $f_{M}$ is an ideal Maxwellian distribution function calculated from moments of $f$, and $\nu$ is the collision frequency. The BGK operator is necessarily conservative in number density, momentum, and energy when $\nu$ is constant with respect to particle velocity as it is in this work. This approximation is appropriate for neutral species, as considered here. For a plasma, $\nu$ is generally known to scale with particle velocity as $v^{-4}$. Assuming constant $\nu$ for a plasma would overestimate energy-fluxes in the high-energy tails of the distribution.

\texttt{Gkeyll} discretizes $f$ on a phase-space grid of up to six dimensions by decomposing $f$ using a set of piecewise polynomials with superlinear order up to $p$ \citep{Arnold2011}. Distribution functions are then evolved in time using a strong-stability-preserving Runge-Kutta method.

\section{Problem Description}
\label{sec:methods}
Distribution functions in this work are 5-dimensional, with two spatial and three velocity space dimensions, $(x, y, v_x, v_y, v_z)$, and have initial conditions derived from hydrostatic equilibrium with 
\begin{equation}
	\boldsymbol{\nabla} p = -nm\boldsymbol{g}.
\end{equation}

All units are normalized using a particle species of mass $m=1.0$, upper bound density $n_1 = 1.0,$ and gravity $g = 1.0$. Initial number density and pressure profiles are as follows

\begin{equation}
	n(y) = \frac{n_0}{2} \tanh\left(\frac{\alpha y}{L_y}\right)+ \frac{3}{2}n_0,
\end{equation}
\begin{equation}
	p(y) = \frac{mgn_0}{2}\left[ \ln \left(\cosh\left(\frac{\alpha y}{L_y}\right)\right) + 3y \right] + \frac{3}{2} n_0 T_0,\\
\end{equation}
where $n_0 = 0.5$ is density at the center of the interface, $L_y = 1.0$ is half the length of the simulation domain in $y$, and $T_0$ is an arbitrary constant chosen to ensure the minimum pressure in the domain is positive. With the density and pressure profiles above, the interface between the high- and low-density regions is continuous and has width defined by $\alpha$. Simulations are initialized with $\alpha=25$ to ensure the width of the interface is small relative to the domain size. This initial density profile corresponds to an Atwood number, $\mathit{A_t} = (n_1 - n_{2})/(n_1+ n_{2})$, of $1/3$. Boundary conditions are periodic in $x$ and static reservoir in $y$, where the edge ghost layers of cells are a continuation of the initial conditions and do not evolve in time. Distribution functions are initially Maxwellian in velocity space, according to, 
\begin{equation}
	f(\boldsymbol{v}) = \frac{n}{(2\pi v_{\mathit{th}}^2)^{3/2}}\exp\left(-\frac{(\boldsymbol{v} - \boldsymbol{u})^2}{2v_{\mathit{th}}^2}\right),
\end{equation}
for initial bulk velocity $\boldsymbol{u}$, where $v_{\mathit{th}} = \sqrt{T/m}$ is thermal velocity. The pressure profile given by Eq. (5) is used to calculate a temperature $T = p/n$, which is then used to initialize the Maxwellian distribution.

While these initial conditions are hydrostatic, they are not a true Boltzmann equilibrium for the case of finite collision frequency, as any deviations from Maxwellian are not immediately damped out by collisions. Additionally, if the collision frequency is not sufficiently high, the interface diffuses and the fluid layers mix before the instability grows. 

\begin{figure}
	\centerline{\includegraphics[width=\linewidth]{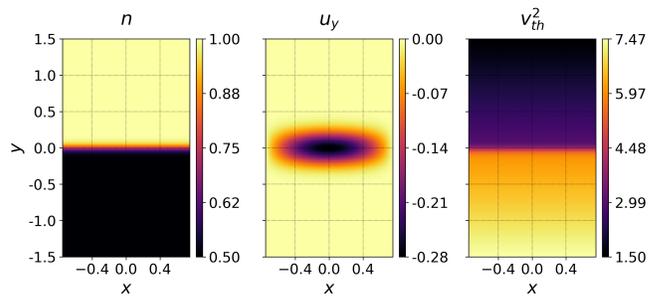}}
	\caption{Initial conditions in number density (left), bulk velocity (center), and square of thermal velocity (right).}
	\label{fig:ic}
\end{figure}

To generate the RT instability, a single-mode sinusoidal perturbation of wavenumber $k$ is applied to the $y$-direction bulk velocity, $u_y$, according to
\begin{equation}
	u_y = -0.1v_{\mathit{th},c}\cos\left(kx\right)\exp\left(-\frac{y^2}{2y_r^2}\right),
\end{equation}
where $k=\pi /(2L_x)$,  $v_{\mathit{th},c}$ is initial thermal velocity at the center of the domain, $L_x = 0.75$ is half the simulation domain length in $x$, and $y_r = L_y/10$ is a characteristic decay length for the perturbation. Initial conditions of $n$, $v_{\mathit{th}}^2$, and $u_y$ are shown in Figure \ref{fig:ic}. 

\section{Results}
\label{sec:results}
\begin{figure}
	\centerline{\includegraphics[width=\linewidth]{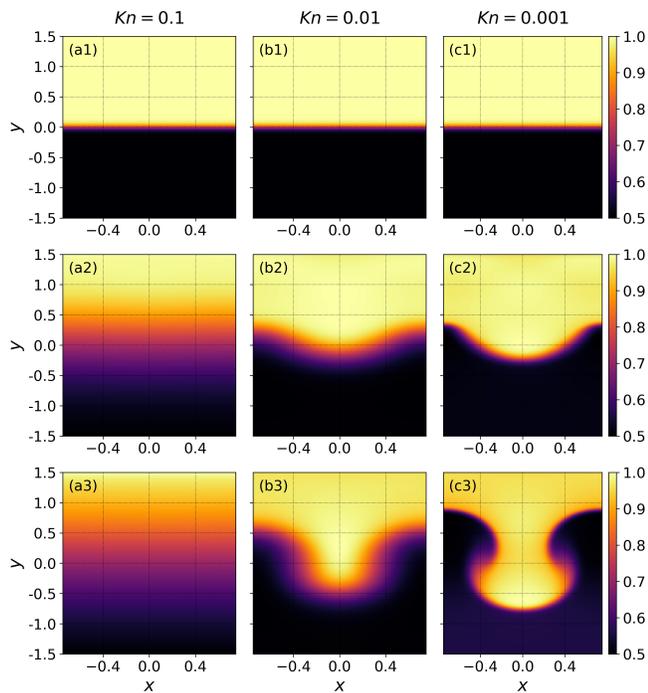}}
	\caption{Time evolution of number density for varying collisionality. Left to right is $\mathit{Kn}$ of 0.1 (a), 0.01 (b), and 0.001 (c). Top to bottom is time 0.0, $1.5\tau_{RT}$, and $3.0\tau_{RT}$.  Note that the low collisionality case (left column) presents no RT instability growth, and the intermediate collisionality case (middle column) presents significantly altered RT instability growth compared to the high collisionality case (right column) which approaches the fluid limit.}
	\label{fig:densevolution}
\end{figure}

\begin{figure}
	\centerline{\includegraphics[width=0.98\linewidth]{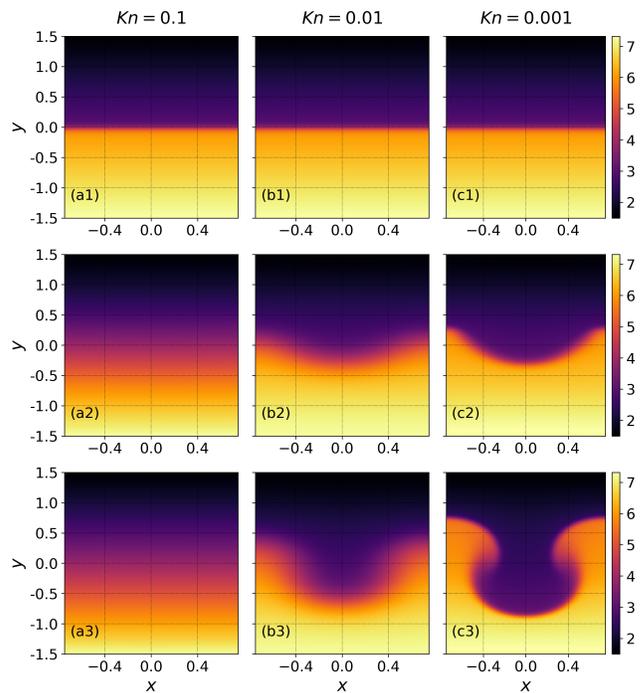}}
	\caption{Time evolution of temperature for varying collisionality. Left to right is $\mathit{Kn}$ of 0.1 (a), 0.01 (b), and 0.001 (c). Top to bottom is time 0.0, $1.5\tau_{RT}$, and $3.0\tau_{RT}$.}
	\label{fig:tempevolution}
\end{figure}

Collision frequencies are calculated from a chosen Knudsen number, $\mathit{Kn} = \lambda_{m}/L_x$, i.e., the ratio between particle mean-free-path $\lambda_m$ and scale length $L_x$. Collision frequencies are assumed to be constant spatially and temporally, according to $\nu = v_{\mathit{th},c}/\lambda_m$. However, collision frequency is generally known to scale with density and temperature \cite{braginskii1965}, and RT instability simulations with spatially-varying collisionality will be explored in future work. In this work, values of $\mathit{Kn}$ are chosen as 0.1, 0.01, and 0.001. Simulations are run to an end time of 3 classical RT instability growth periods, $\tau_{RT} = 1/\sqrt{kgA_t}$. Time evolution of number density and temperature for each case is shown in Figure \ref{fig:densevolution} and Figure \ref{fig:tempevolution} respectively, for times 0, $1.5\,\tau_{RT}$, and $3.0\,\tau_{RT}$. 

The fluid interface diffuses in all cases due to finite collisionality. As mean-free-path increases from the limit of infinite collisionality, particles stream past one another over longer distances without interacting. The net result is a mixing of the fluid layers that speeds up as mean-free-path increases, as particles are not affected by the pressure gradient until a collision event. With no perturbation, the interface continues to diffuse until the fluid layers mix completely. 

For the lowest collisionality case, the interface diffuses so quickly relative to the RT instability growth time scale that there is effectively no interface where the instability can form. As collisionality increases by an order of magnitude, the interface diffuses slowly enough that the RT instability is able to grow. At the end time, the expected bubble and spike structures are present with diffuse edges. The most collisional case approaches the expected fluid result, with minimal diffusion of the interface and mushroom structures on the bubble and spike as secondary Kelvin-Helmholtz instabilities form. The temperature distribution exhibits identical behavior to the density evolution. Note that the growth of the RT instability for the intermediate case is slower than that of the highly collisional case. 

\begin{figure}
	\centerline{\includegraphics[width=0.8\linewidth]{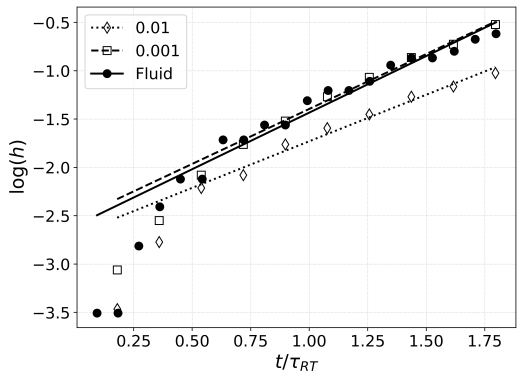}}
	\caption{Logarithm of $h$, the difference between spike and bubble heights, as a function of time for $\mathit{Kn} = 0.01 \text{ and } 0.001$ and a fluid simulation using the Euler equations. Data points early in time are excluded from the fit due to dynamic diffusion of the interface for the kinetic cases and wave launching for the fluid case. Note as $\mathit{Kn}$ decreases, the RT instability growth rate approaches the fluid simulation result.}
	\label{fig:growth}
\end{figure}

In order to quantify the effects of collisionality on RT instability growth, an approach similar to \cite{Sagert2015} and \cite{Duff1962} is used to calculate a growth rate, $\gamma_{0}$, that includes viscous and diffusive effects, 
\begin{equation}
    \gamma_{0} = \sqrt{kg\mathit{At}+ \nu_v^2k^4} - (\nu_v + \xi)k^2,
\end{equation}
where $\nu_v = v_{\mathit{th},c}\lambda_m/2$ is the kinematic viscosity, and $\xi = \nu_v$ is the diffusion coefficient. Note that \cite{Sagert2015} and \cite{Duff1962} include an additional factor for dynamic diffusion effects to calculate a time-dependent growth rate, which has been neglected here. Because the primary dynamic diffusion effect is the diffusion of the interface, which occurs exclusively early in the simulation, those early data points are excluded from the growth rate calculation to achieve a constant linear growth rate that describes RT instability growth for the majority of the simulation. Growth rates are calculated using $h$, the difference between the top of the bubble and the bottom of the spike, and are presented in Figure \ref{fig:growth} for the case of $\mathit{Kn}$ = 0.01 and 0.001, compared with a neutral fluid simulation using the Euler equations. It is assumed that kinetic simulations converge to those of the Euler equations in the limit of infinite collisionality as non-ideal transport becomes negligible. Early data points are also ignored for the fluid simulations, as the perturbation to $u_y$ causes waves to be launched that interfere with RT instability growth early in time. Growth rates calculated from the linear fits in Figure \ref{fig:growth} are compared with theoretical growth rates in Table \ref{tab:growth}. 
\begin{table}[b]
\caption{\label{tab:growth}%
Values of RT instability growth rates, calculated from simulation ($\gamma$) and theory ($\gamma_0$).}
\begin{ruledtabular}
\begin{tabular}{ccc}
\textrm{Case}&
\textrm{$\gamma$}&
\textrm{$\gamma_0$}\\
\colrule
       $\mathit{Kn} = $ 0.01   & 0.9635 & 0.9723 \\
       $\mathit{Kn} = $ 0.001   & 1.1369 & 1.1603\\
       Fluid & 1.1708 & 1.1816\\
\end{tabular}
\end{ruledtabular}
\end{table}
There is good agreement between the calculated growth rates and the theoretical growth rates with static diffusion, and as $\mathit{Kn}$ increases, $\gamma$ and $\gamma_0$ approach the fluid result. The slight decrease in agreement from the 0.01 $\mathit{Kn}$ case to the 0.001 case is likely due either to the presence of diffusion in the kinetic case or not capturing the transition from time-varying growth to linear growth as well in the data output frames (i.e., the transition is between data points 3 and 4 for the 0.001 case).  While fluid simulations of the RT instability have been performed with viscosity \citep{menikoff1977unstable, srinivasan2014mitigating, song2020survey, bera2021effect}, the presence of a fluid viscosity alone is insufficient to explain the diffusion of the interface seen here (including the dynamic diffusion effects early in time).  To explain the kinetic parameter regime of the intermediate collisionality case, this work is a first to probe into a detailed kinetic interpretation of the RT instability.

While highly collisional regimes asymptoting to fluid results are reasonably well-understood for the neutral fluid RT instability, intermediate collisional regimes require kinetic simulations since the fluid model is no longer valid in these regimes. Variation in RT instability growth as a function of collisionality implies the emergence of kinetic effects as collisionality decreases and distribution functions are allowed to become less Maxwellian. A metric to quantify non-Maxwellian distributions spatially can aid in probing the 5-dimensional distribution function by highlighting potential areas of variation from equilibrium. In an attempt to capture the spatial distribution of such variations, a density analogue is constructed from the distribution function and a constructed Maxwellian as follows,

\begin{equation}
	n_{N}(\boldsymbol{x}) = \int |f(\boldsymbol{x},\boldsymbol{v}) - f_M(\boldsymbol{x},\boldsymbol{v})| \text{d}^3\boldsymbol{v}.
\end{equation}
 
This non-Maxwellian density allows for spatial representation of non-Maxwellian distribution functions and has units of density, allowing for simple comparison to the density profiles in Figure \ref{fig:densevolution}. Non-Maxwellian density for each $\mathit{Kn}$ is presented below in Figure \ref{fig:nm}. Note that recent work by \citet{Cagas2021} shows that a boundary layer forms at reservoir boundaries for the Vlasov-BGK model, Eqs. (1) and (2). The boundary layer is approximately one mean-free-path wide and non-Maxwellian. Therefore, three layers of cells at the top and bottom of the domain are omitted in Figure \ref{fig:nm} in order to maintain a useful color scale for the regions of interest. As expected, high collisionality leads to a decrease in $n_N$ by approximately an order of magnitude between the most and least collisional cases. For the case where the RT instability does not develop, $n_N$ simply follows an almost identical distribution to density, comparing Figure \ref{fig:densevolution} (a3) to the center plot of Figure \ref{fig:nm}. However, for the cases where the RT instability develops, the interfaces appear as regions of peak $n_N$. Magnitudes of $n_N$ are small relative to $n$, even for the least collisional case that has the highest peak $n_N$. 

\begin{figure}
	\centerline{\includegraphics[width=\linewidth]{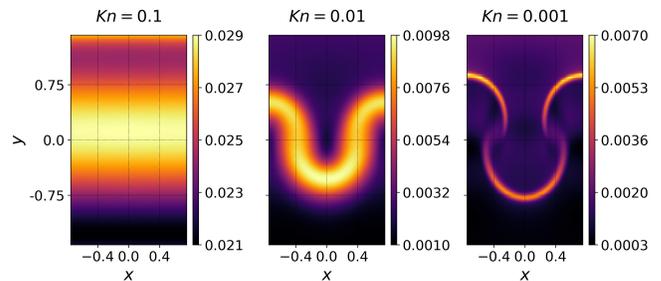}}
	\caption{Density of non-ideal distribution, according to Eq. (9), for $\mathit{Kn}$ of 0.1 (left), 0.01 (center), and 0.001 (right), normalized to number density. Note the varying color scale for each subplot.}
	\label{fig:nm}
\end{figure}

\begin{figure*}[h]
	\centerline{\includegraphics[width=0.85\linewidth]{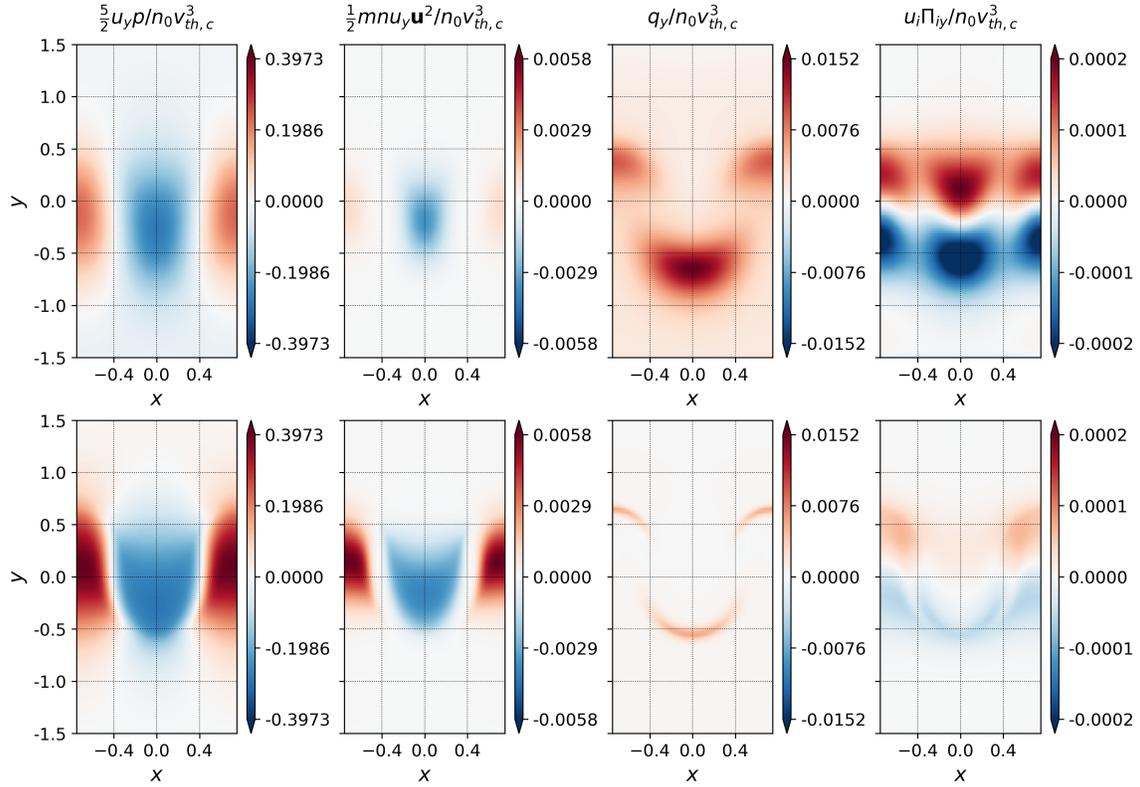}}
	\caption{Terms of the expanded particle energy-flux (3.5) in the $y$-direction for $\mathit{Kn}$ of 0.01 (top) and 0.001 (bottom), normalized to $n_0v_{th,c}^3$. Energy-flux is calculated at normalized time 3.0$\tau_{\mathit{RT}}$ for the 0.01 case and 2.1$\tau_{\mathit{RT}}$ for the 0.001 case to have similar amplitudes. Note the varying color scale of each column.}
	\label{fig:ideal}
\end{figure*}

To further characterize the effect of varying collisionality, two higher moments of the distribution function are defined,
\begin{equation}
    \mathcal{P}_{\mathit{ij}} = m\int v_i v_j f \text{d}^3\boldsymbol{v},
\end{equation}
\begin{equation}
	\mathcal{Q}_{\mathit{ijk}} = m\int v_i v_j v_k f\text{d}^3 \boldsymbol{v}.
\end{equation}
As in \citet{Wang2015}, by defining $w_i = v_i - u_i$, Eq. (11) can be expanded and tensor contracted to get the particle energy-flux (using Einstein's summation convention),
\begin{equation}
    \frac{1}{2}\mathcal{Q}_{\mathit{iik}} = \underbrace{\frac{5}{2}u_kp + \frac{1}{2}mnu_k\boldsymbol{u}^2}_{\text{I}} + \underbrace{q_k + u_i\Pi_{\mathit{ik}}}_{\text{II}},
\end{equation}
where
\begin{equation}
    q_k = \frac{1}{2}m\int w_iw_iw_kf\text{d}^3\boldsymbol{v},
\end{equation}
is the heat flux vector in the gas frame, and the stress tensor $\Pi_{\mathit{ij}}$ is related to the pressure tensor,
\begin{equation}
    P_{\mathit{ij}} = m\int w_iw_jf\text{d}^3\boldsymbol{v},
\end{equation}
by $\Pi_{\mathit{ij}} = P_{\mathit{ij}} - p\delta_{\mathit{ij}}$ with scalar pressure $p = P_{\mathit{ii}}/3$. The pressure tensor is also related to the second moment by $\mathcal{P}_{\mathit{ij}} = P_{\mathit{ij}} + mnu_iu_j$. Note that the use of collision frequency that is independent of particle velocity leads to an overestimation of energy fluxes in the high-energy tails of the distribution if charged species are considered instead of neutral species.  In the case of charged species, the energy-fluxes presented here will be greater in magnitude than those calculated with a collision frequency that varies with velocity.  This work considers neutral species. Individual terms are grouped in Eq. (12) by whether they arise from Maxwellian parts of the distribution (group I), or non-Maxwellian parts (group II). Group I will be referred to as ideal terms, while group II are the non-ideal terms. The $y$-component of each term of Eq. (12) (normalized to $n_0v_{\mathit{th},c}^3$) is plotted in Figure \ref{fig:ideal} for the cases of $\mathit{Kn} = 0.01$ and 0.001 for times with similar instability amplitude. Note that magnitudes of each column (term) vary by orders of magnitude, so color scales are distinct by column to show spatial features. The first ideal term is the dominant term by several orders of magnitude at its peak for both cases. As collisionality increases, all terms increase in magnitude, though the ideal terms increase more than the non-ideal terms. This can be seen by taking the ratio of the average of the absolute values of the ideal terms to that of the non-ideal terms. The ratio is 21.5 for the less collisional case and 283.8 for the more collisional case, indicating the particle energy-flux becomes less dominated by the ideal terms as collisionality decreases.  This is an important and impactful result as it is the first to present an order of magnitude increase in the importance of the non-ideal terms for the less collisional (more kinetic) case of the RT instability.  The overall increase in energy-flux with increased collisionality, even when comparing similar amplitudes of RT instability growth, relates to the increase in growth rate shown in Table \ref{tab:growth}, as larger total flux leads to faster instability growth.

By taking moments of a first-order Chapman-Enskog expansion of the BGK collision operator, expressions for the heat flux, $q_{i,\text{BGK}}$, and stress tensor $\Pi_{ij,\text{BGK}}$, can be obtained assuming a nearly Maxwellian distribution,

\begin{gather}
    \Pi_{ij,\text{BGK}} = -\frac{p}{\nu}\left(\frac{\partial u_i}{\partial x_j} \frac{\partial u_j}{\partial x_i} - \frac{2}{3}\frac{\partial u_k}{\partial x_k}\delta_{ij} \right), \\
    q_{i,\text{BGK}} = -\frac{5p}{2m\nu}\frac{\partial T}{\partial x_i}.
\end{gather}

\begin{figure}
    \centering
    \subfloat{ \includegraphics[width=0.8\linewidth]{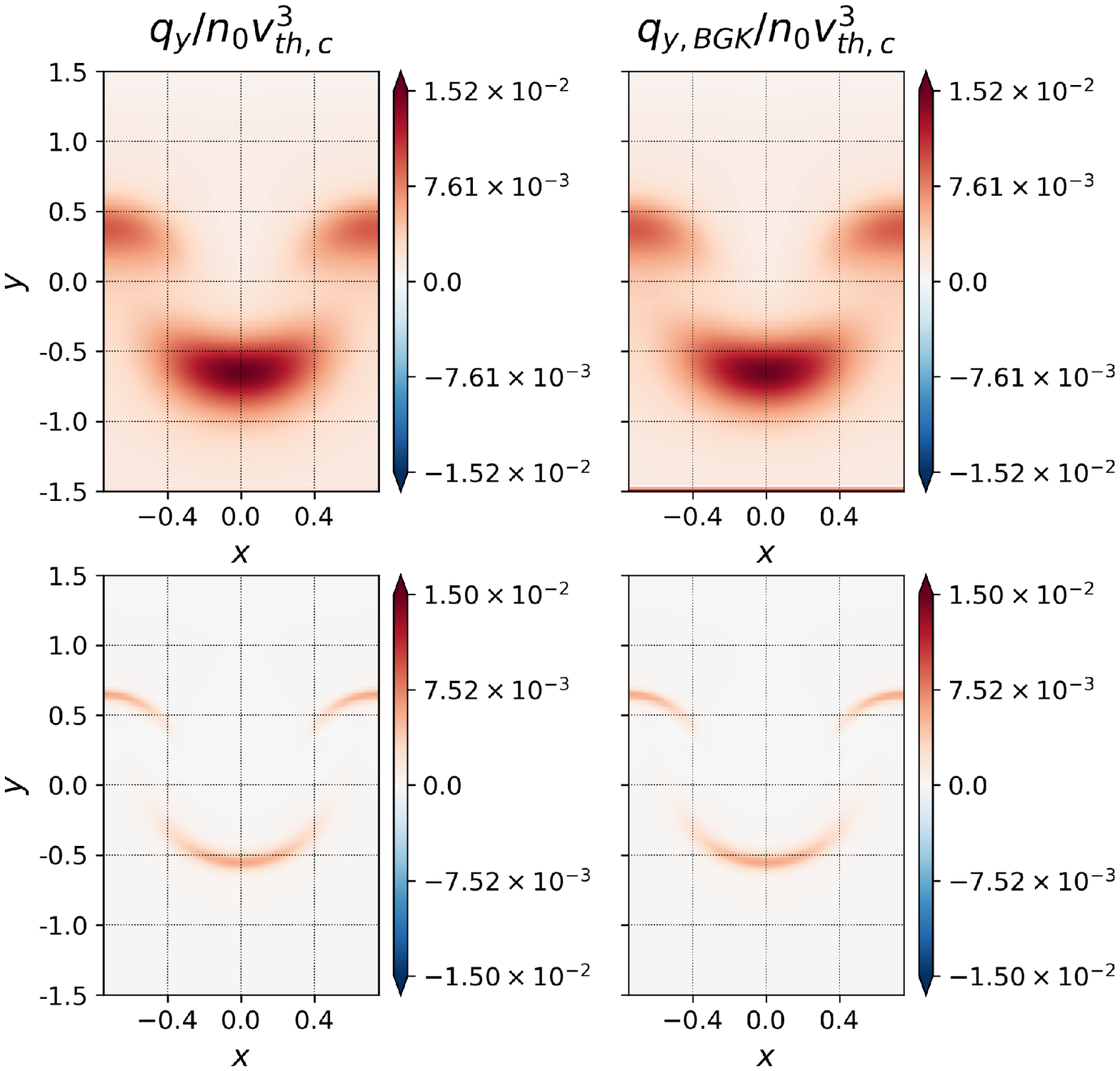}}
    \hfill
    \subfloat{ \includegraphics[width=0.8\linewidth]{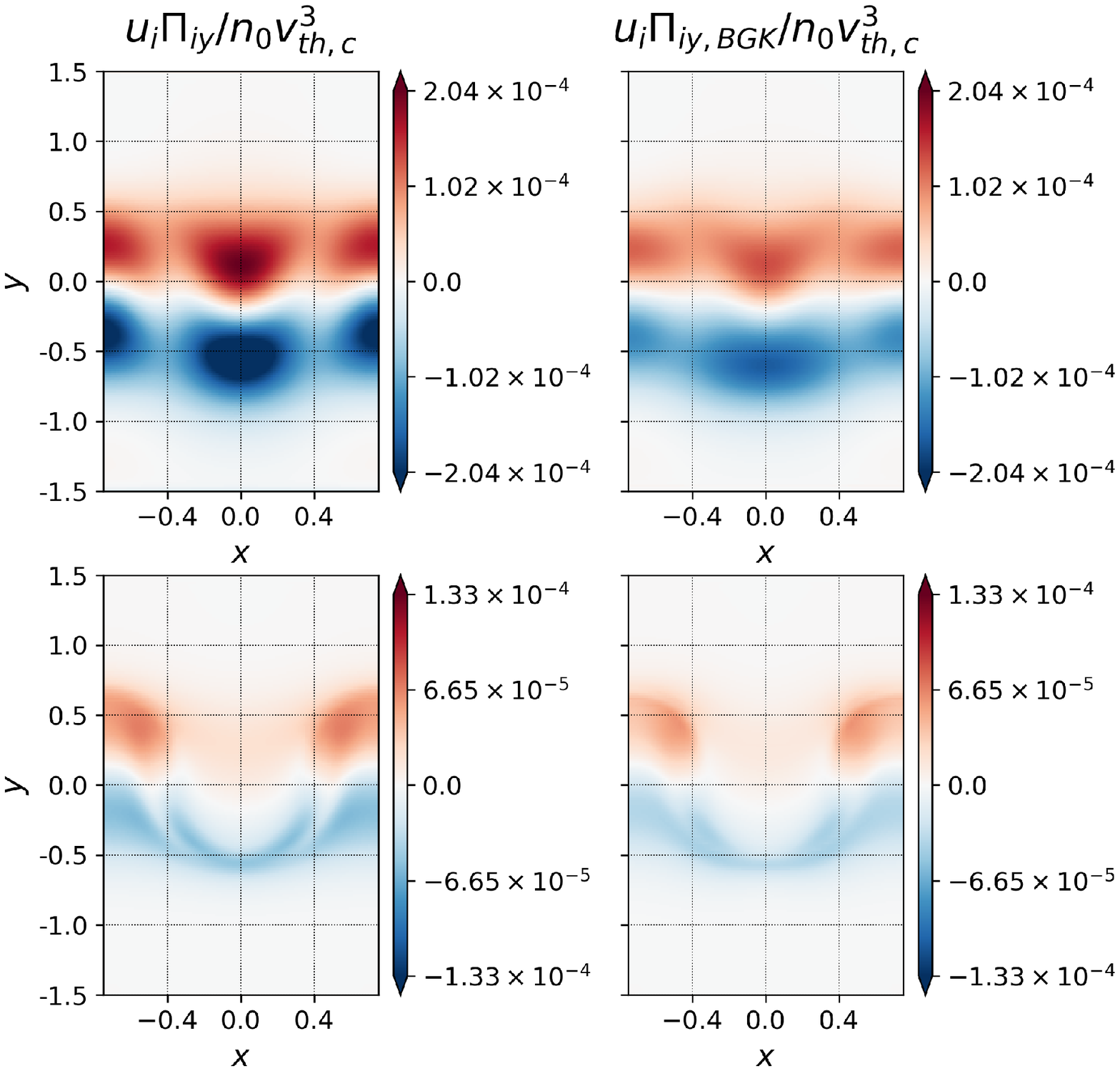}}

    \caption{Comparison of energy-flux non-ideal terms, Eq. (12), calculated from distribution function and those calculated from a first-order Chapman-Enskog expansion of the collision operator. Top and bottom rows of each comparison are $\mathit{Kn}$ = 0.01 and 0.001, respectively. Note the similarities in spatial distribution and magnitude and that color scales are constant by term and row (collisionality). Stress terms show more discrepancy because they are calculated from several stress tensor elements, so errors compound.}
    \label{fig:BGK}
\end{figure}

Figure \ref{fig:BGK} presents the non-ideal terms of the particle energy-flux calculated directly from the distribution function with those calculated from the expansion. Note color scale is held constant for each term compared across the two different values of collisionality. For both degrees of collisionality, the heat flux terms are similar in both magnitude and spatial distribution. The stress terms show more deviation between true and approximate results likely due to the fact that two stress tensor elements are involved in the calculation, so errors from the first-order approximation compound. As collisionality decreases from $\mathit{Kn}$ = 0.001 to 0.01, the approximate stress term deviates more from the direct calculation because the assumption of near-equilibrium distribution becomes less accurate with decreasing collisionality.

\begin{figure}
    \centering
    \includegraphics[width=\linewidth]{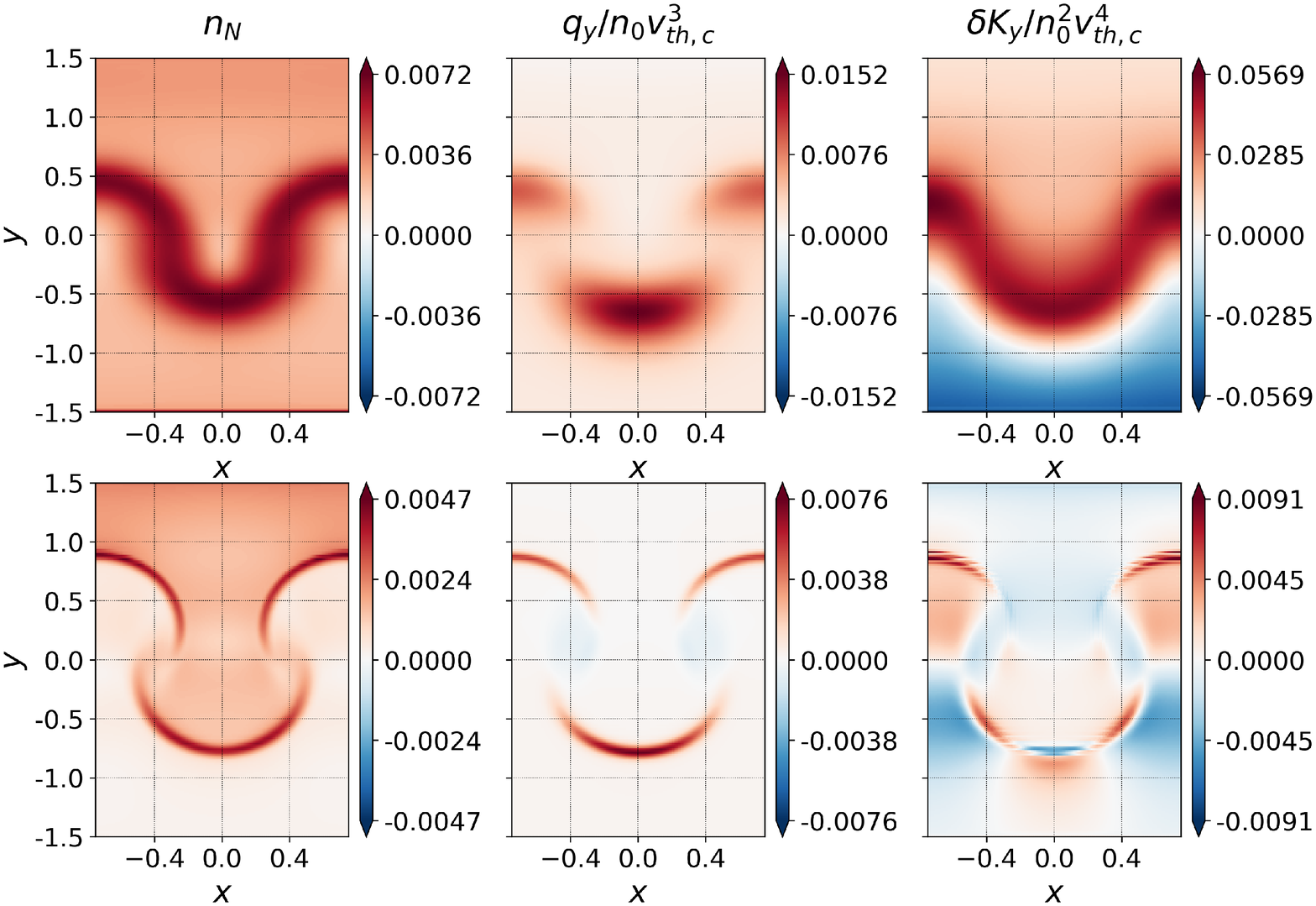}
    \caption{Non-Maxwellian density, $n_N$, compared with gas-frame $y$-direction skewness, $q_y$, and excess kurtosis, $\delta K_y$. Note the presence of local extrema for all quantities around the RT instability interface.}
    \label{fig:higher}
\end{figure}

Higher moments of the distribution function are also measures of non-ideal distribution, so the spatial distribution of gas-frame higher moments should correlate with $n_N$. Presented in Figure \ref{fig:higher} are comparisons of $n_N$, $y$-direction vector skewness, $q_y$, and $y$-direction excess kurtosis,
\begin{equation}
    \delta K_y = \int w_i^4 f \text{d}^3\boldsymbol{v} - \int w_i^4 f_M \text{d}^3\boldsymbol{v}.
\end{equation}
As expected, the distribution of $n_N$ aligns with those of the higher moments. Additionally, the magnitudes of the normalized higher moments increase as collisionality decreases, which is expected as decreased collisions deviate from a Maxwellian distribution function towards a more kinetic regime.   The evolution of the intermediate collisionality RT instability is clearly distinguished from the high collisionality regime to explain the kinetic effects that produce the difference in growth rates and morphology.  These are the first results to present a high-fidelity kinetic interpretation of the classical RT instability in low and intermediate collisionality regimes where fluid models are inadequate. 

\section{\label{sec:conclusion}Conclusion}
Single-mode Rayleigh-Taylor instabilities are successfully simulated in 2x3v using the continuum-kinetic capabilities of \texttt{Gkeyll} for a range of collisionalities. As mean-free-paths become smaller relative to the width of the simulation domain, the resulting instability approaches the classical fluid result, as expected. Growth rates estimated using static viscosity and diffusion agree well with calculated growth rates when early dynamic diffusion of the interface is left out of the fit. Non-Maxwellian density, the velocity space integration of the difference between a local particle distribution function and a corresponding Maxwellian distribution calculated from the first three moments, shows that, as collisionality increases, the distribution function approaches a Maxwellian (fluid) distribution. Local maxima in non-Maxwellian density also occur around the primary areas of transport, i.e., the edges of the bubble and spike. A decomposition of the particle energy-flux shows that transport is dominated by terms that arise from the Maxwellian parts of the distribution, and the ideal terms of the expansion become more dominant as collisionality increases toward the fluid limit.

An important and novel contribution of this work is in the intermediate collisional cases that are not accessible with traditional fluid models and require kinetic modeling.  The continuum-kinetic model used in this work provides unique access to the full noise-free distribution function to investigate the kinetic regime.  Simulations of intermediate collisional cases show significantly altered RT instability evolution compared to the high collisionality fluid-like cases highlighting the importance of kinetic physics through higher moments of the distribution function.  These higher moments include the heat flux vector, which is the third moment indicating the skewness of the distribution, and the fourth moment indicating the kurtosis of the distribution.  The heat flux vector plays a more significant role relative to inertial terms in the intermediate collisional cases compared to the highly collisional cases. A quantitative comparison shows an order of magnitude difference in the ratio of the non-ideal terms to the ideal terms when comparing the intermediate collisional cases to the highly collisional fluid-like cases.  These kinetic effects are primarily noted in the region of the RT instability interface.  Regimes of intermediate collisionality often occur in astrophysical and laboratory plasmas requiring a kinetic model due to the invalidity of the fluid model for these cases, highlighting the significance and relevance of the results presented here.

\begin{acknowledgments}
This work was supported by the National Science Foundation CAREER award under grant number PHY-1847905. A. Hakim is supported through the U.S. Department of Energy contract No. DE-AC02-09CH11466 for the Princeton Plasma Physics Laboratory. The authors acknowledge Advanced Research Computing at Virginia Tech for providing computational resources and technical support that have contributed to the results reported within this paper. URL: \url{http://www.arc.vt.edu}
\end{acknowledgments}

\appendix
\section{Getting Gkeyll and reproducing results}
Readers may reproduce our results and also use Gkeyll for their applications. The code and input files used here are available online. Full installation instructions for
Gkeyll are provided on the Gkeyll website \citep{gkylDocs}. The code can be installed on Unix-like operating systems (including Mac OS and Windows using the Windows Subsystem for Linux) either by installing the pre-built binaries using the conda package manager (\url{https://www.anaconda.com}) or building the code via sources. The input files used here are under version control and can be obtained from the repository at \url{https://github.com/ammarhakim/gkyl-paper-inp/tree/master/2022_PRE_RayleighTaylor}

% The \nocite command causes all entries in a bibliography to be printed out
% whether or not they are actually referenced in the text. This is appropriate
% for the sample file to show the different styles of references, but authors
% most likely will not want to use it.
% \nocite{*}

%\bibliography{ref}% Produces the bibliography via BibTeX.
%apsrev4-2.bst 2019-01-14 (MD) hand-edited version of apsrev4-1.bst
%Control: key (0)
%Control: author (8) initials jnrlst
%Control: editor formatted (1) identically to author
%Control: production of article title (0) allowed
%Control: page (0) single
%Control: year (1) truncated
%Control: production of eprint (0) enabled
%

\end{document}